\documentclass[aps,prl,reprint,citeautoscript,superscriptaddress,showpacs,floatfix, longbibliography]{revtex4-1}

\usepackage{graphicx,epsfig,float}
\usepackage{amssymb}
\usepackage{amsmath}
\usepackage[varg]{txfonts}
\usepackage{listings}
\usepackage{xcolor}
\usepackage{soul}

\definecolor{codegreen}{rgb}{0,0.6,0}
\definecolor{codegray}{rgb}{0.5,0.5,0.5}
\definecolor{codepurple}{rgb}{0.58,0,0.82}
\definecolor{backcolour}{rgb}{0.95,0.95,0.92}

\lstdefinestyle{mystyle}{
    backgroundcolor=\color{backcolour},   
    commentstyle=\color{codegreen},
    keywordstyle=\color{magenta},
    numberstyle=\tiny\color{codegray},
    stringstyle=\color{codepurple},
    basicstyle=\ttfamily\footnotesize,
    breakatwhitespace=false,         
    breaklines=true,                 
    captionpos=b,                    
    keepspaces=true,                 
    numbers=left,                    
    numbersep=5pt,                  
    showspaces=false,                
    showstringspaces=false,
    showtabs=false,                  
    tabsize=2
}

\allowdisplaybreaks
\begin{document}
\title{Modular implementation of the linear and cubic-scaling orbital minimization methods in electronic structure codes using atomic orbitals}

\author{Irina V. Lebedeva}
\email{liv\_ira@hotmail.com}
\affiliation{CIC nanoGUNE BRTA, 20018 Donostia-San Sebasti\'an, Spain}
\affiliation{Catalan Institute of Nanoscience and Nanotechnology - ICN2 (CSIC and BIST), Campus UAB, 08193 Bellaterra, Spain}
\affiliation{Simune Atomistics, Avenida de Tolosa 76, 20018 Donostia-San Sebasti\'an, Spain}
\author{Alberto Garc\'ia}
\email{albertog@icmab.es}
\affiliation{Institut de Ci\`encia de Materials de Barcelona (ICMAB-CSIC), 08193 Bellaterra, Spain}
\author{Emilio Artacho}
\email{ea245@cam.ac.uk}
\affiliation{CIC nanoGUNE BRTA, 20018 Donostia-San Sebasti\'an, Spain}
\affiliation{Donostia International Physics Center DIPC, 20018 Donostia-San Sebasti\'an, Spain}
\affiliation{Theory of Condensed Matter, Cavendish Laboratory, University of Cambridge, Cambridge CB3 0HE, UK}
\affiliation{Ikerbasque, Basque Foundation for Science, 48011 Bilbao, Spain}
\author{Pablo Ordej\'on}
\email{pablo.ordejon@icn2.cat}
\affiliation{Catalan Institute of Nanoscience and Nanotechnology - ICN2 (CSIC and BIST), Campus UAB, 08193 Bellaterra, Spain}

\begin{abstract}
We present a code modularization approach to design efficient and massively parallel cubic and linear-scaling solvers for electronic structure calculations using atomic orbitals. The modular implementation of the orbital minimization method, in which linear algebra and parallelization issues are handled via external libraries, is demonstrated in the SIESTA code. The DBCSR and ScaLAPACK libraries are used for algebraic operations with sparse and dense matrices, respectively. The MatrixSwitch and libOMM libraries, recently developed within the Electronic Structure Library, facilitate switching between different matrix formats and implement the energy minimization. We show results comparing the performance of several cubic-scaling algorithms, and also demonstrate the parallel performance of the linear-scaling solvers, and their supremacy over the cubic-scaling solvers for insulating systems with sizes of several hundreds of atoms.
\end{abstract}

\maketitle

\section{Introduction}
The success of electronic structure theory \cite{Martin2004} 
in modeling new materials and devices \cite{Mardirossian2017,Krylov2018} has stimulated the development of hundreds of electronic structure codes \cite{MOLSSI}. Historically almost all of these individual software packages are written in distinct ways, although many tasks performed are similar.
  Except for numerical and performance related libraries such as basic 
linear algebra subroutines (BLAS) \cite{BLAS}, higher-level
linear algebra utilities (serial Linear Algebra PACKage - LAPACK \cite{LAPACK} and its parallel counterpart, Scalable LAPACK - ScaLAPACK \cite{SCALAPACK}), message passing interface (MPI) level \cite{MPI}, etc., significant parts of the codes are replicated with some variations. 
Electronic structure packages are growing rapidly incorporating more and more new features. Also the codes have to adapt to the constant hardware evolution, which in the case of monolithic code architecture implies significant efforts on re-engineering. In this situation, it seems more efficient to change the traditional monolithic paradigm of 
software development to the modular one in which common tasks 
arise \cite{Oliveira2020}. 

In addition, such an approach allows to separate the tasks related to high-level routines focused on the calculation of physical properties from the implementation of the underlying routines for parallelization and algebra. On one hand, this means that the implementation of new models and algorithms becomes much simpler and does not require the knowledge of technical details related to parallelization. On the other hand, much better performance is achieved using specialized external libraries and thus much larger systems can be modeled. Significant efforts (e. g., the European project MAX\cite{MAX}) are underway to stimulate the paradigm change in software design and to facilitate exascale computing.
Here we show the benefits of modularization by the example of SIESTA \cite{SIESTA,Ordejon1996,Soler2002,Sanchez-Portal1997,Garcia2020}.

SIESTA \cite{SIESTA,Ordejon1996,Soler2002,Sanchez-Portal1997,Garcia2020} was specifically designed for linear-scaling calculations\cite{Ordejon1993,Ordejon1995} in which the computational time grows linearly with the number of atoms
\cite{Ordejon1993,Ordejon1995,Mauri1994,Galli1996,Goedecker1999,Bowler2012}. Such methods make possible calculations of large systems at a considerably less computational cost compared to
common cubic-scaling approaches.
SIESTA uses strictly localized atomic-like functions for basis sets in which the Hamiltonian and overlap matrices, $\mathbf{H}$ and $\mathbf{S}$, are sparse. If, additionally, the confinement of the wavefunctions is imposed, the coefficient matrix $\mathbf{C}$ expanding wavefunctions in the basis is also sparse.
Reducing the problem of
solving the Kohn-Sham equations to the minimization of a properly constructed energy functional within the Ordej\'on-Mauri  \cite{Ordejon1993,Ordejon1995,Mauri1994, Mauri1993} and Kim \cite{Kim1995} approaches, the inversion of the overlap matrix is avoided and only expressions involving products and sums of sparse $\mathbf{H}$, $\mathbf{S}$ and $\mathbf{C}$ matrices need to be computed, 
all in linear-scaling effort.

\begin{figure}
   \centering
 \includegraphics[width=0.8\columnwidth]{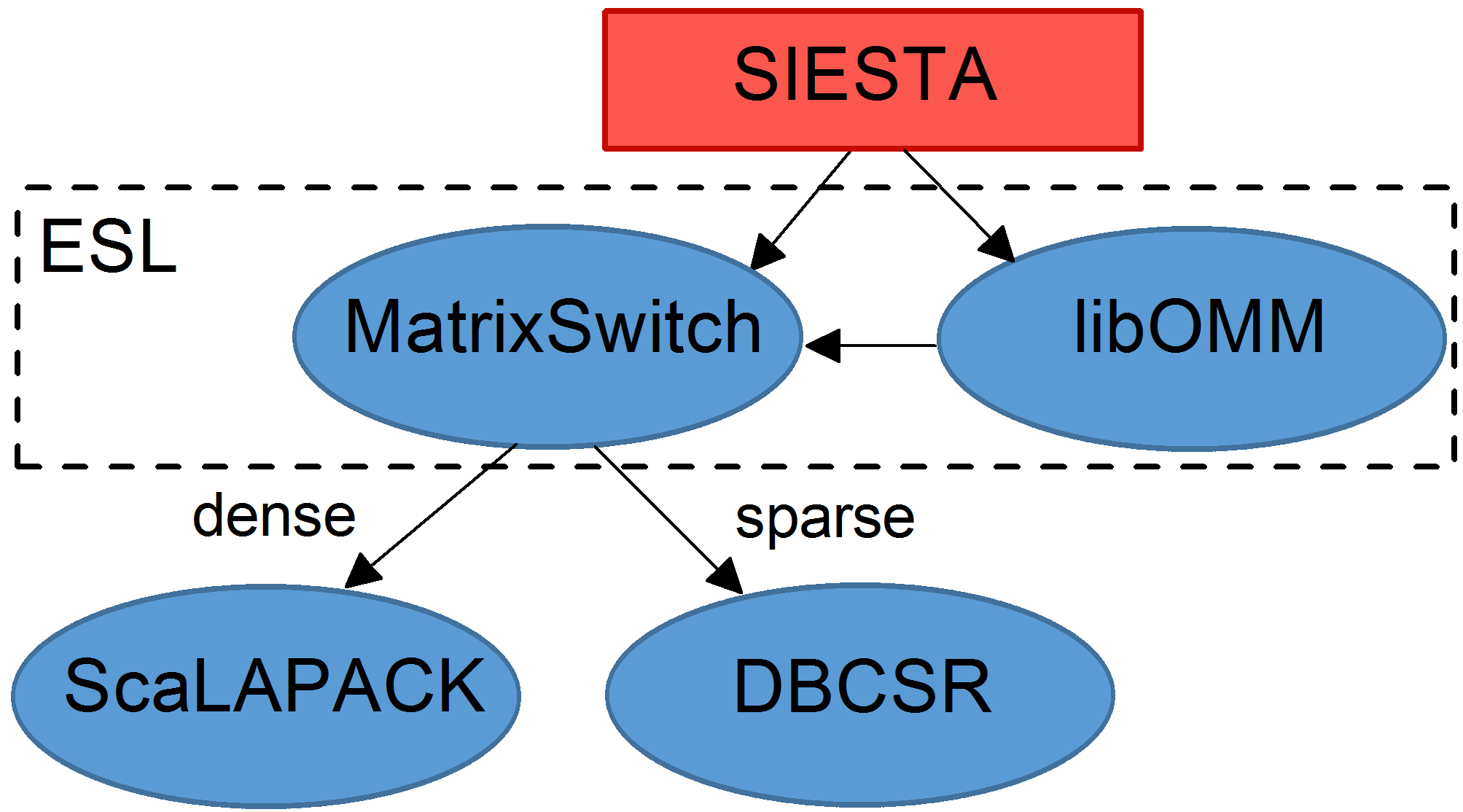}
   \caption{The use of libraries within the revised OMM solver in the electronic structure code SIESTA \cite{SIESTA,Ordejon1996,Soler2002,Sanchez-Portal1997,Garcia2020}. The red rectangular box corresponds to SIESTA. Blue ellipses indicate the libraries used \cite{DBCSR,SCALAPACK,LIBOMM,MS,MSDBCSR,Oliveira2020,Borstnik2014,Corsetti2014}. The libraries in the dashed frame belong to the Electronic Structure Library (ESL) \cite{ESL,Oliveira2020}. The arrows demonstrate calls to the libraries. } 
   \label{fig:code_scheme}
\end{figure}

The linear-scaling solvers in SIESTA, although available from the start
\cite{Ordejon1996}, 
are not widely used in practice. One of the reasons is that the implementation of these physical methods involved also coding of the algebra and parallelization of sparse matrices, which inevitably increased the code complexity and hindered progress. Recent efforts on 
linear-scaling methods have produced the distributed block compressed sparse row (DBCSR) library \cite{DBCSR} that efficiently handles algebraic operations for sparse matrices and is massively parallelized \cite{Borstnik2014,CP2K}. Using this library, we have implemented an improved and more reliable version of linear-scaling solvers in SIESTA (Fig. \ref{fig:code_scheme}).

Another recent initiative that has helped re-designing SIESTA is the Electronic Structure Library \cite{ESL,Oliveira2020}, a collaboration platform for shared software development.
We use ESL's libOMM library \cite{LIBOMM,Oliveira2020} distributed within the omm-bundle  \cite{omm-bundle}. 
  It encodes the Ordej\'on-Mauri \cite{Ordejon1993,Ordejon1995,Mauri1994, Mauri1993} 
and Kim \cite{Kim1995} functionals, originally without the additional approximation of
wave-function confinement, rendering dense $\mathbf{C}$ matrices and cubic scaling.
 Such an approach provides an alternative to conventional cubic-scaling methods, which can be 
faster in long simulations by avoiding computationally expensive orthonormalization and using history on previous steps \cite{Corsetti2014}. 
  We refer to unconstrained minimization methods of suitable energy functionals, 
with either linear or cubic scaling, as the orbital minimization method (OMM) \cite{Tsuchida2007,Bowler2010, Bowler2012,Corsetti2014}. In  libOMM \cite{LIBOMM,Corsetti2014,Oliveira2020}, the minimization is customarily performed 
via conjugate gradients (CG). The parameters of the quartic function describing the energy dependence along the search direction are computed analytically \cite{Ordejon1995,Corsetti2014}.

Although the original libOMM library provides cubic scaling \cite{LIBOMM,Corsetti2014,Oliveira2020}, it has been straightforward to extend it to linear scaling, the equations being almost the same, the key difference being the use of sparse matrices instead of dense. 
Normally two separate pieces of the code dealing with sparse and dense matrices would be used for the same equations. This code duplication can be avoided in libOMM thanks to the MatrixSwitch (MS) library \cite{MS}, an interface between high-level physical routines and low-level routines for matrix algebra. MS, which is also distributed within the omm-bundle \cite{omm-bundle} of ESL \cite{ESL,Oliveira2020}, simplifies the coding of matrix operations and allows a single code independent of matrix format, by means of format-independent high-level commands. Depending on matrix format, MS calls the appropriate linear algebra library. An example of calculations using the MS library is shown in Listing \ref{lst:MS_example} (see Supplementary Material for MS overview).   

Recently the MS library was extended to support sparse matrices \cite{MSDBCSR,Oliveira2020} via the DBCSR \cite{DBCSR,Borstnik2014,CP2K} library. Here we consider dense and sparse matrices in the {\it pddbc} (parallel-distributed dense block cyclic) and {\it  pdcsr}  (parallel-distributed compressed sparse row) MS formats for which algebraic operations are handled with the help of the ScaLAPACK \cite{SCALAPACK} and DBCSR \cite{DBCSR} libraries, respectively. Although basic functionality for sparse matrices was already provided in this recent MS version \cite{MSDBCSR,Oliveira2020}, a revision of the library was needed towards treating sparse and dense matrices on the same footing, and to enable linear-scaling calculations. The incorporation of the solver library into an electronic structure code also implies additional matrix manipulations such as conversions between the matrix formats supported by the code and the solver library as well as reading and writing of restart files. The corresponding subroutines have been here implemented in MS and are discussed below.

After a brief overview of the OMM approaches, the new implementation of linear and cubic-scaling OMM in SIESTA is presented, including the necessary changes in the MS and libOMM libraries forming part of ESL. The results of the first tests are discussed, and recommendations on the efficient use of OMM are given.

\begin{lstlisting}[language={[90]Fortran}, caption={An example of the calculation of the total charge in the OMM approach using the MatrixSwitch library.},label={lst:MS_example}]
! Calculation of the total charge
! n_e=2Tr[(2I-CdSC)CdSC], where I is the identity
! matrix, S is the overlap matrix in the basis of 
! localized atomic orbitals, C is the 
! coefficient matrix that describes the expansion 
! of the wavefunctions in the basis of localized
! atomic orbitals and Cd is the Hermitian
! conjugate of matrix C. The matrix format is
! determined by the variable m_storage. The Cd
! and S matrices are provided as an input.

! Allocating CdS matrix of type m_storage and size
! n x m (the number of wavefunctions x the basis
! set size) consisting of blocks of size b_n x b_m
call m_allocate(CdS,n,m,label=m_storage,blocksize1=b_n,blocksize2=b_m)
! Allocating the SW matrix
call m_allocate(SW,n,n,label=m_storage,blocksize1=b_n,blocksize2=b_n)
! Allocating the I matrix
call m_allocate(I,n,n,label=m_storage,blocksize1=b_n,blocksize2=b_n)
! Allocating the auxiliary matrix A
call m_allocate(A,n,n,label=m_storage,blocksize1=b_n,blocksize2=b_n)
! Calculating CdS
call mm_multiply(Cd,'n',S,'n',CdS,1.0_dp,0.0_dp)
! Calculating SW = CdSC
call mm_multiply(CdS,'n',Cd,'c',SW,1.0_dp,0.0_dp)
! Setting the identity matrix I
call m_set(I,'a',0.0_dp,1.0_dp)
call m_add(SW,'n',I,-2.0_dp,4.0_dp) ! I=4I-2SW
call mm_trace(I,SW,n_e) ! n_e=Tr[I x SW]
call m_deallocate(CdS) ! Deallocating the CdS matrix
call m_deallocate(SW) ! Deallocating the SW matrix
call m_deallocate(I) ! Deallocating the I matrix
call m_deallocate(A) ! Deallocating the A matrix
\end{lstlisting}

\section{Overview of OMM approaches}
In density functional theory (DFT) \cite{Hohenberg1964,Kohn1965}, the problem of finding the ground state of a many-electron system is reduced to an energy minimization for the system of $2n$ non-interacting electrons moving in an effective potential and described by one-particle states \{$|\psi_i\rangle$\} ($i=1,...,n$) each of which is occupied by two electrons of opposite spin (assuming no spin polarization, for simplicity). 
The set of states \{$|\psi_i\rangle$\} is one of the many possible bases in the occupied subspace of the Hilbert space of the system and can be chosen orthonormal or not. In the latter case \cite{Artacho1991}, the overlap matrix $\mathbf{S_W}$ with the elements  $(S_\mathrm{W})_{ij}=\langle\psi_i|\psi_j\rangle$ is not the identity matrix ($\mathbf{S_W}\ne \mathbf{I}$, $(I_\mathrm{W})_{ij}=\delta_{ij}$) and the density matrix operator that determines the projection onto the occupied subspace is then given by  
 \begin{equation} \label{eq_rho}
\hat{\rho}=2\sum_{i,j=1}^n|\psi_i\rangle\left(\mathbf{S_W}^{-1}\right)_{ij}\langle\psi_j|,
\end{equation}
involving the inverse of $\mathbf{S_W}$. Here and below we limit our consideration to insulating systems \cite{Ordejon1995,Mauri1994, Mauri1993} . The linear-scaling methods applicable to metals are discussed, e.g., in Refs. \cite{Goedecker1994, Goedecker1995, Corkill1996, Mohr2018}.

The corresponding band structure energy becomes \cite{Ordejon1995,Ordejon1993,Mauri1994, Mauri1993}  
\begin{equation} \label{eq_en}
E=\mathrm{Tr}\left[\hat{H}\hat{\rho}\right]=2\mathrm{Tr}\left[\mathbf{S_W}^{-1}\mathbf{H_W}\right],
\end{equation}
where $\hat{H}$ is the Hamiltonian operator and $\mathbf{H_W}$ is the matrix with the elements  $(H_\mathrm{W})_{ij}=\langle\psi_i|\hat{H}|\psi_j\rangle$. 
Note that the traces in Eq. \eqref{eq_en} are taken on spaces of different dimensions: the size of the basis set for the first, and of the occupied states in the second. Also, the second equality holds for zero temperature.

In the basis of $m$ functions \{$|\phi_i\rangle$\} (strictly localized atomic orbitals in SIESTA)
 \begin{equation} \label{eq_psi}
|\psi_i\rangle=\sum_{\mu=1}^m C_i^\mu|\phi_\mu\rangle,
\end{equation}
where we refer to $\mathbf{C}$ as the coefficient matrix. Then  $\mathbf{H_W}=\mathbf{C}^\dagger\mathbf{H}\mathbf{C}$ and $\mathbf{S_W}=\mathbf{C}^\dagger\mathbf{S}\mathbf{C}$, where $H_{ij}=\langle\phi_i|\hat{H}|\phi_j\rangle$, $S_{ij}=\langle\phi_i|\phi_j\rangle$ and
 $\mathbf{C}^\dagger$ is the Hermitian conjugate of $\mathbf{C}$.
The energy functional in Eq. (\ref{eq_en}) is minimized to find the ground state energy. The most common approach is direct diagonalization of the Hamiltonian matrix $\mathbf{H}$ (an $m\times m$ matrix for the basis set of size $m$). Energy and charge density are then obtained using the wavefunctions and energies of the $n$ lowest eigenstates.  In contrast, in the iterative approaches \cite{Payne1992}, the energy is minimized with respect to variations in the states \{$|\psi_i\rangle$\}. Here one needs to calculate the inverse of the overlap matrix $\mathbf{S_W}^{-1}$ or impose the orthonormality condition $(S_\mathrm{W})_{ij}=\delta_{ij}$. In any case, the computational time increases as $O(n^3)$ with the system size, while the memory required to store the wavefunctions grows as $O(n^2)$.

In OMM approaches \cite{Ordejon1993,Ordejon1995,Mauri1994, Mauri1993}, the expensive orthonormalization step is avoided via the modification of the energy functional in such a way that it automatically induces the orthonormalization of the wavefunctions during minimization:
\begin{equation} \label{eq_OM}
\begin{split}
\tilde E&= 2\mathrm{Tr}\left[(\mathbf{I_W}+(\mathbf{I_W}-\mathbf{S_W}))\mathbf{H_W}\right]\\
&=2\mathrm{Tr}\left[(2\mathbf{I_W}-\mathbf{C}^\dagger\mathbf{S}\mathbf{C})\mathbf{C}^\dagger\mathbf{H}\mathbf{C}\right].
\end{split}
\end{equation}
This expression can be derived from consideration of Lagrange multipliers \cite{Ordejon1993,Ordejon1995} or expansion of the inverse overlap matrix to first order in the deviation from the identity \cite{Mauri1993,Mauri1994}: $\mathbf{S_W}^{-1} \approx \mathbf{I_W} + (\mathbf{I_W} - \mathbf{S_W})$. The solution obtained from Eq. (\ref{eq_OM}) is the same as from Eq. (\ref{eq_en}).  

Within the same approximation, the density matrix of Eq.~\eqref{eq_rho} is computed as  \cite{Ordejon1995} 
 \begin{equation} \label{eq_rho_OM}
 \boldsymbol{\rho}=\mathbf{C}\left(\mathbf{I_W}+(\mathbf{I_W}-\mathbf{S_W})\right)\mathbf{C}^\dagger=2\mathbf{C}(2\mathbf{I_W}-\mathbf{C}^\dagger\mathbf{S}\mathbf{C})\mathbf{C}^\dagger
\end{equation}
and the forces on atom $I$ as \cite{Ordejon1995} 
 \begin{equation} \label{eq_force}
\mathbf{F}_I=-\mathrm{Tr}\left[\boldsymbol{\rho}\frac{\partial \mathbf{H}}{\partial \mathbf{R}_I}\right]+\mathrm{Tr}\left[\boldsymbol{\rho}_E\frac{\partial \mathbf{S}}{\partial \mathbf{R}_I}\right],
\end{equation}
where we refer to $\boldsymbol{\rho}_E=2\mathbf{C}\mathbf{H_W}\mathbf{C}^\dagger$ as the ``energy density''.

If the basis functions and wavefunctions are chosen to be strictly localized, the Hamiltonian, overlap and coefficient matrices, $\mathbf{H}$, $\mathbf{S}$ and $\mathbf{C}$, are sparse and $O(n)$ scaling with system size is achieved \cite{Ordejon1993,Ordejon1995,Mauri1994, Mauri1993}. Note that this is not the case for Eqs.~\eqref{eq_rho} and \eqref{eq_en} as the inverse of $\mathbf{S}$ is not sparse (although sub-cubic scaling can be achieved using selected inversion to compute just the needed elements of the inverse \cite{Lin2011}). In the case of periodic systems, localized wavefunctions are close to the Wannier functions that decay exponentially with the distance from the center of localization in insulators and in metals at a finite temperature. Imposing localization constraints on the wavefunctions, however, leads to a deviation from the exact solution of Eqs.~\eqref{eq_en} and \eqref{eq_OM}. Also the localized wavefunctions obtained are not strictly orthonormal and do not comply with the system symmetries \cite{Kim1995}. However, the degree of approximation can be controlled with the cutoff radius $R_\mathrm{C}$ for the wavefunctions. Here we limit our consideration to insulators with a substantial band gap, where $R_\mathrm{C}$ of several \AA~is normally enough \cite{Ordejon1995,Mauri1994}.

In the Ordej\'on-Mauri functional \cite{Ordejon1993,Ordejon1995,Mauri1994, Mauri1993}, the localization of the wavefunctions gives rise to many shallow local minima and flat regions in which the algorithm can be trapped for a long time during the energy minimization. This problem is solved in the Kim functional \cite{Kim1995} by including unoccupied states and introducing a chemical potential $\eta$, i.e. the energy separating occupied and unoccupied states. The corresponding functional is obtained by (1) an eigenspectrum shift $\mathbf{H}\to\mathbf{H}-\eta \mathbf{S}$, (2) changing dimensions of $\mathbf{C}$ from $m \times n$ to $m \times n^\prime$, where $n^\prime > n$, and (3) changing the energy functional in Eq. (\ref{eq_OM})  as $\tilde E\to\tilde E+\eta n$, and energy density $\boldsymbol{\rho}_E$ in Eq. (\ref{eq_force}) as $\boldsymbol{\rho}_E\to\boldsymbol{\rho}_E+\eta\boldsymbol{\rho}$. It should be noted, however, that although the multiple-minima problem is solved in the Kim functional, it is sometimes hard to choose a proper value for $\eta$. It should always lie within the band gap, but the bands can move up and down during self-consistency or molecular dynamics (MD), $\eta$ possibly getting into the valence or conduction bands and, as a result, converging to an erroneous solution. Care should be taken to ensure that the solution reproduces the correct number $2n$ of electrons. 

If the localization constraints on the wavefunctions are removed, the exact solution of Eqs. (\ref{eq_en}) and (\ref{eq_OM}) is obtained \cite{Corsetti2014}. Even in this case, however, one energy minimization can demand many CG iterations. This relates to the problem of length-scale or kinetic energy ill-conditioning \cite{Payne1992,Bowler1998}. The efficiency of the CG algorithm depends on the ratio of the maximal and minimal extremal curvatures of the function minimized, which in OMM are determined by the maximal and minimal eigenvalues of the Hamiltonian. The eigenspectrum of the Hamiltonian is broad given the large kinetic energy of high-energy eigenstates. Although such states contribute negligibly to the ground-state solution, the problem becomes ill-conditioned and the convergence is slow. It is, however, possible effectively to reduce the width of the eigenspectrum by suppressing the kinetic energy contribution of high-energy states through preconditioning \cite{Gan2001,Mostofi2003}, by which the CG gradient matrix is multiplied by the preconditioning matrix \cite{Corsetti2014}:
\begin{equation} \label{eq_precon}
\mathbf{P}=\left(\mathbf{S}+\frac{\mathbf{T}}{\tau_\mathrm{T}}\right)^{-1}, 
\end{equation}
where ${\tau}_\mathrm{T}$ is the scale for kinetic energy preconditioning and $\mathbf{T}$ is the kinetic energy matrix. Another approach for improving the efficiency of CG minimizations is reducing the generalized eigenvalue problem to the standard form via the Cholesky factorization \cite{Corsetti2014}. Both of these approaches involve matrices that are not sparse (the preconditioning matrix or the reduced Hamiltonian) and are considered here only for cubic-scaling OMM.

\begin{figure}
   \centering
 \includegraphics[width=\columnwidth]{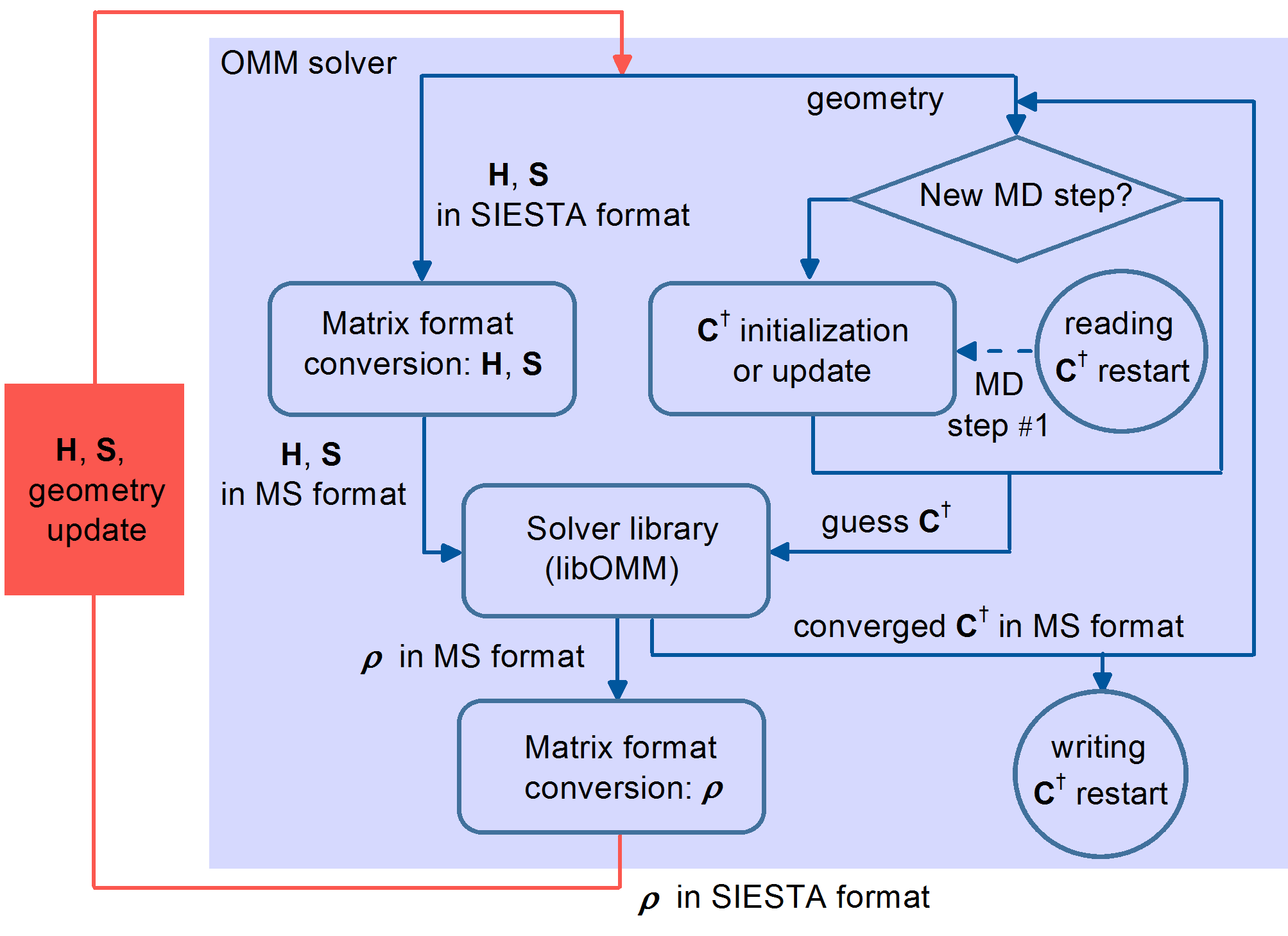}
   \caption{Scheme of the revised OMM solver. Blocks within the solver are shown in blue. The rest of the SIESTA code is shown as a red block. Arrows indicate data flow. Hamiltonian, overlap and density matrices are denoted as $\mathbf{H}$, $\mathbf{S}$ and $\boldsymbol{\rho}$. The Hermitian conjugate of the coefficient matrix of expansion of the wavefunctions in the basis of localized atomic  orbitals is denoted as $\mathbf{C}^\dagger$. The Hamiltonian and overlap matrices are converted from internal SIESTA to MatrixSwitch (MS) format for further calculation of $\boldsymbol{\rho}$ and $\mathbf{C}^\dagger$ with the help of the libOMM library. The restart file for $\mathbf{C}^\dagger$ can be read once at the first molecular dynamics (MD) step.} 
   \label{fig:solver_scheme}
\end{figure}

\section{Modular solver architecture}
\subsection{Solver input and output} 
A scheme of the implemented OMM solver is shown in Fig. \ref{fig:solver_scheme}. At each self-consistent-field (SCF) step, the solver receives as an input the Hamiltonian and overlap matrices in the basis of strictly localized atomic orbitals, $\mathbf{H}$ and $\mathbf{S}$, and the information on the system geometry. SIESTA uses for matrices the standard compressed sparse row format, that is the matrix information is stored in local one-dimensional (1D) arrays containing data values and column indices for individual nonzero elements of local rows as well as indices of the first nonzero elements and numbers of nonzero elements for each local row. The blocks of rows are distributed on a 1D process grid (Fig. \ref{fig:m_copy}a). Here and in MS we refer to this format as {\it pdrow} to distinguish from the {\it pdcsr} format supported by DBCSR. $\mathbf{H}$ and $\mathbf{S}$ are received by the solver in the {\it pdrow} format. The density matrix $\boldsymbol{\rho}$ is the output, also in {\it pdrow} (see Eq. (\ref{eq_rho_OM})). This matrix is used to update $\mathbf{H}$ for the next SCF step outside the solver. At the end of each MD step, the solver is called again to compute the energy density matrix $\boldsymbol{\rho}_E$, which, along with  $\boldsymbol{\rho}$, is later used to calculate forces (see Eq. (\ref{eq_force})) and stresses. The scheme of the $\boldsymbol{\rho}_E$ calculation is analogous to that of $\boldsymbol{\rho}$ shown in Fig. \ref{fig:solver_scheme}.

\subsection{Solver library} 
The solver uses the libOMM library \cite{LIBOMM,Corsetti2014,Oliveira2020,omm-bundle} to perform the CG minimization of the energy functional given by Eq. (\ref{eq_OM}). As an input, the libOMM library requires $\mathbf{H}$ and $\mathbf{S}$, as well as the initial guess for $\mathbf{C}^\dagger$, in one of the MS formats \cite{MS,MSDBCSR,Oliveira2020,omm-bundle}. As an output, it provides the converged  $\mathbf{C}^\dagger$, and $\boldsymbol{\rho}$ or  $\boldsymbol{\rho}_E$ in the same format. The {\it pddbc} format is used for parallel calculations with dense matrices. In this case, all matrix elements are stored and algebraic operations are performed using the ScaLAPACK library \cite{SCALAPACK}. The matrix is divided into 2D blocks distributed on a 2D  or 1D process grid. For parallel calculations with sparse matrices, the {\it pdcsr} format is used. The matrix is also divided into 2D blocks distributed on a 1D or 2D process grid (Figs. \ref{fig:m_copy}b and c, respectively). However, in this case, zero blocks are not stored. The algebraic operations are performed by the DBCSR library \cite{DBCSR,Borstnik2014,CP2K}. At the moment, libOMM supports only equal rectangular blocks. 

The equations implemented in the libOMM library are compatible with all OMM flavours discussed in the previous section, including the Ordej\'on-Mauri
and Kim functionals, with and without localization constraints.
However, to make the libOMM library functional for sparse matrices, some parts to the code have been reformulated. Now block-size information is passed to the MS library during the allocation of intermediate matrices required for the CG minimization using \texttt{m\_allocate()} (see Supplementary Material). Also sparsity is imposed on the gradient matrix $\mathbf{G}$ (with the elements $G^\mu_i=\partial \tilde E/\partial (C^{\mu}_i)^\mathrm{*}$) following the sparsity pattern of the initial guess for $\mathbf{C}$.  Already during $\mathbf{G}$ calculation \cite{Corsetti2014}, only matrix elements that fit into the sparsity pattern are computed in the contributions to $\mathbf{G}$ that are given by products of matrices (using \texttt{keep\_sparsity = true} option of \texttt{mm\_multiply()}). In the rest of the contributions, nonzero elements that do not fit into the sparsity pattern are omitted and no longer stored, while zero elements within the sparsity pattern are stored as zeros. The sparsity of the density ($\boldsymbol{\rho}$) and energy density ($\boldsymbol{\rho}_E$) matrices is assumed to be the same as of the overlap matrix $\mathbf{S}$ and only elements of these matrices that fit into the sparsity pattern are computed. Additionally, the expression for the calculation of $\boldsymbol{\rho}_E$ has been corrected as compared to the previous libOMM version \cite{Corsetti2014} in accordance with Eqs. (\ref{eq_OM}) -- (\ref{eq_force}) and Ref. \cite{Ordejon1995}. The Cholesky factorization and kinetic energy preconditioning are available only for dense matrices.

\begin{figure}
   \centering
 \includegraphics[width=\columnwidth]{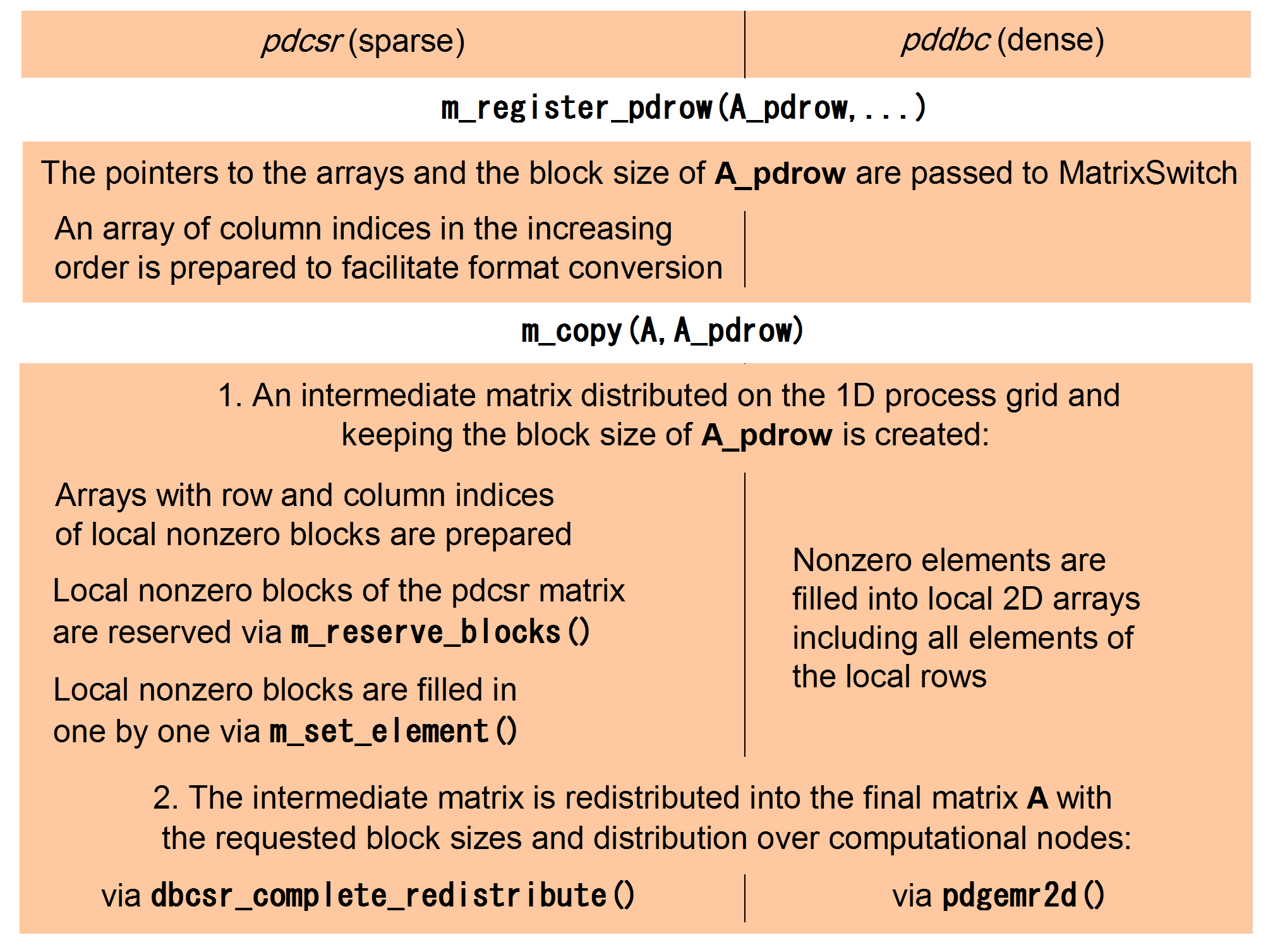}
   \caption{A series of calls to the MatrixSwitch library required for the format conversion of matrix $\mathbf{A}$ from the {\it pdrow} format used in SIESTA ({\bf A\_pdrow}) to the {\it pdcsr} and {\it pddbc} MatrixSwitch formats ({\bf A}) handled with the DBCSR and ScaLAPACK libraries, respectively.} 
   \label{fig:format_conversion}
\end{figure}

\begin{figure*}
   \centering
 \includegraphics[width=\textwidth]{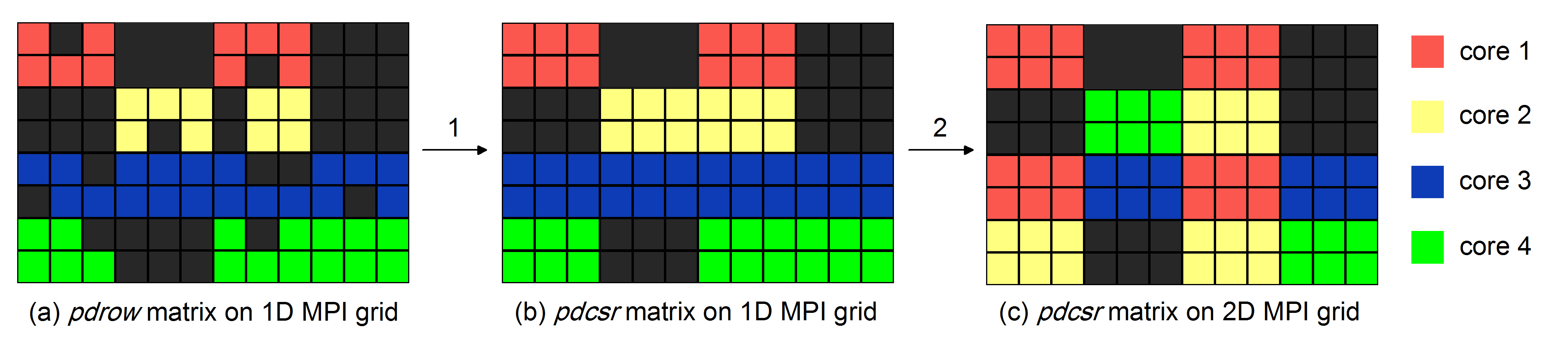}
   \caption{Example of matrix format conversion from the {\it pdrow} format used in SIESTA to {\it pdcsr} MatrixSwitch format handled with DBCSR: (a) {\it pdrow} matrix distributed on the 1D process grid with 4 CPU cores, (b) {\it pdcsr} matrix with $2 \times 3$ blocks distributed on the same 1D process grid and (c) {\it pdcsr} matrix with $2 \times 3$ blocks distributed on the  $2 \times 2$ 2D process grid.  Arrows indicate steps 1 and 2 of subroutine \texttt{m\_copy()} as explained in Fig. \ref{fig:format_conversion}. Small squares represent elements of the $8\times12$ matrix. Black squares are zero elements that are not stored. Red, yellow, blue and green squares correspond to elements stored on cores 1, 2, 3 and 4, respectively.}
   \label{fig:m_copy}
\end{figure*}

\subsection{$\mathbf{C}^\dagger$ matrix format conversion} 
 In order to incorporate the libOMM library into SIESTA within the OMM solver, the following steps are required (see Fig. \ref{fig:solver_scheme}): (1) matrix format conversion from/to the SIESTA format to/from the MS formats and (2) initialization and update of $\mathbf{C}^\dagger$, according to the current geometry of the system. The matrix format conversion is realized using calls to MS subroutines \texttt{m\_register\_pdrow()} and \texttt{m\_copy()} (see Supplementary Material). The  first of this subroutines has been added to the MS library and the second one has been extended to allow the conversion from/to the {\it pdrow} format to/from the {\it pdcsr} and {\it pddbc} formats. The conversion is performed as follows (Fig. \ref{fig:format_conversion}). First the pointers to arrays of the {\it pdrow} matrix and its block size are passed to MS. Then a {\it pdcsr}/{\it pddbc} matrix distributed on the 1D process grid with the same block size for rows as the initial {\it pdrow} matrix is filled in element by element (Fig. \ref{fig:m_copy}). The missing elements of the {\it pddbc} matrix or within nonzero blocks of the {\it pdcsr} matrix are filled with zeros. Note that to speed up the conversion and guarantee linear scaling, column and row indices of nonzero blocks of the {\it pdcsr} matrix should be passed to the DBCSR library before filling the values via the call to \texttt{m\_reserve\_blocks()} (see Supplementary Material). Once the 1D-distributed {\it pdcsr}/{\it pddbc} matrix is ready, it can be redistributed on a 2D process grid. In the case when the final matrix is distributed on the 1D process grid and has the same block size for rows at the initial  {\it pdrow} matrix, the last step is omitted.

The conversion from {\it pdcsr} and {\it pddbc} to {\it pdrow} is implemented in a  similar way. It is assumed that the row and column indices of nonzero elements of the  {\it pdrow} matrix are already known. Only the values of the matrix elements are restored.

\subsection{$\mathbf{C}^\dagger$ matrix initialization and update} 
The initialization of the $\mathbf{C}^\dagger$ matrix in the sparse form is performed in SIESTA in the following way. It is supposed that each atom carries the number of localized wavefunctions equal to the atomic charge (in units of elementary charge) divided by two, $Q_\mathrm{at}/2$. If $Q_\mathrm{at}$ is odd, $(Q_\mathrm{at}+1)/2$ localized wavefunctions are assigned to one atom and  $(Q_\mathrm{at}-1)/2$ to the next one. This procedure is repeated for all the atoms in the system. Then the $\mathbf{C}^\dagger$ matrix in the  {\it pdrow} format with the total number of rows that corresponds to the total number of localized wavefunctions, $N_\mathrm{WF} = Q/2$, where $Q$ is the sum of atomic charges in the system, is prepared. The local rows are assigned according to the block size $b_\mathrm{WF}$. By default, it equals the block size for the basis functions, $b_\mathrm{BF}$, multiplied by the ratio of the total number $N_\mathrm{WF}$ of localized wavefunctions to the basis set size $N_\mathrm{BF}$: $b_\mathrm{WF}=b_\mathrm{BF}N_\mathrm{WF}/N_\mathrm{BF}$. For each local row, the local environment of the atom hosting the corresponding localized wavefunction is analyzed. The row elements that correspond to atoms beyond some cutoff radius $R_\mathrm{C}$ from the atom considered are supposed to be zero. The row elements that correspond to atoms within the cutoff radius $R_\mathrm{C}$ are initialized by random values. This sparsity pattern is maintained during the energy functional minimization. The $\mathbf{C}^\dagger$ matrix in the {\it pdrow} format is converted to the {\it pdcsr} or {\it pddbc} formats in the same manner as the Hamiltonian and overlap matrices, $\mathbf{H}$ and $\mathbf{S}$. 

It should be mentioned also that the initial cutoff radius $R_\mathrm{C,ini}$ for initialization of the $\mathbf{C}^\dagger$ matrix can be set different from $R_\mathrm{C}$ used for the energy minimization. Choosing a small initial radius $R_\mathrm{C,ini}$ (several \AA) helps to avoid convergence problems and is useful not only in calculations with sparse matrices but also with dense ones.

At each new MD step, the sparsity pattern of the $\mathbf{C}^\dagger$ matrix is checked again. The elements that now should be zero because the corresponding atoms got away by more than $R_\mathrm{C}$ are set to zero and no longer stored. The elements corresponding to the atoms that got closer than $R_\mathrm{C}$ are now stored and treated as nonzero but are assigned to zero as the initial guess.  Linear extrapolation of the $\mathbf{C}^\dagger$ matrix based on the information from the two previous MD steps is also possible. 

\subsection{$\mathbf{C}^\dagger$ matrix input and output} 
The restart file for the $\mathbf{C}^\dagger$ matrix can be written at each SCF step  and read at the beginning of the run. These operations are performed by calling new MS subroutines \texttt{m\_read()} and \texttt{m\_write()}, respectively (see Supplementary Material). If the $\mathbf{C}^\dagger$ matrix in the {\it pdcsr} or {\it pddbc} format is distributed on a 2D process grid, it is first converted into a 1D-distributed matrix (by analogy with the format conversion routines). Then the blocks of rows are consecutively passed to the head core and written to the file. To read the file, the reverse operations are performed. The block sizes and process grid for the $\mathbf{C}^\dagger$ matrix do not need to be the same as used when writing the restart information. Upon reading, the sparsity pattern of the $\mathbf{C}^\dagger$ matrix is corrected according to the current system geometry.

\subsection{SIESTA input parameters} 

The input parameters for SIESTA corresponding to the revised OMM solver are described in Table \ref{table:input}. To use the OMM solver, \texttt{SolutionMethod} should be set to \texttt{BLOMM} (OMM with block matrices). 
  
\begin{table*}
    \caption{Principal input parameters for the revised OMM solver in SIESTA (\texttt{SolutionMethod BLOMM}) and their default values.}
   \renewcommand{\arraystretch}{1.2}
   \setlength{\tabcolsep}{12pt}
    \resizebox{\textwidth}{!}{
        \begin{tabular}{lll}
\hline
Input parameter & Default value & Description    \\\hline
\texttt{OMM.UseSparse} & \texttt{true} & Whether to use sparse matrices \\
\texttt{OMM.UseKimFunctional}  & \texttt{true} & Whether to use the Kim \cite{Kim1995} (or Ordej\'on-Mauri \cite{Ordejon1993,Ordejon1995,Mauri1994, Mauri1993}) functional \\
\texttt{OMM.Use2D} & \texttt{true} & Whether to distribute matrices on a 2D process grid \\
\texttt{OMM.ReadCoeffs} & \texttt{false} & Whether to read the initial localized wavefunctions (LWFs), i.e. the $\mathbf{C}^\dagger$   \\
 & & matrix, from the restart file (*.WF\_COEFFS\_BLOMM) \\
\texttt{OMM.WriteCoeffs} & \texttt{false} & Whether to write the LWFs ($\mathbf{C}^\dagger$ matrix) to the restart file \\
\texttt{OMM.RelTol} & \texttt{$10^{-9}$} & The tolerance for the energy convergence in conjugate-gradient (CG) iterations.  \\
 & &  When $2(E_n-E_{n-1})/(E_n+E_{n-1})$, where $E_n$ is the energy at CG iteration $n$,   \\& &  becomes smaller than this tolerance, CG iterations are stopped \\
\texttt{OMM.BlockSizeC} & $b_\mathrm{WF}=b_\mathrm{BF}N_\mathrm{WF}/N_\mathrm{BF}$ & The block size for LWFs (rows of the $\mathbf{C}^\dagger$ matrix). By default, equals the block  \\
 &  &  size for the basis functions $b_\mathrm{BF}$ (input parameter \texttt{BlockSize}) multiplied by the  \\
 &  & ratio of the total number $N_\mathrm{WF}$ of LWFs to the basis set size $N_\mathrm{BF}$ \\
\texttt{OMM.Eta} & \texttt{0 eV} & The chemical potential for the Kim functional \\
\texttt{OMM.RcLWF} & \texttt{9.5 Bohr} & The cutoff radius $R_\mathrm{C}$ for LWFs determining the sparsity pattern of the $\mathbf{C}^\dagger$ matrix \\
\texttt{OMM.RcLWFInit} & \texttt{0 Bohr} & The initial cutoff radius $R_\mathrm{C,ini}$ for LWFs. It is the same as \texttt{OMM.RcLWF} if set to \texttt{0} \\
\texttt{OMM.Extrapolate} & \texttt{false} & Whether to estimate LWFs at the next molecular dynamics (MD) step by \\
 & & the linear extrapolation of the results of two last MD steps \\\hline
\multicolumn{3}{l}{\it Only for the cubic-scaling OMM}\\\hline
\texttt{OMM.Precon} & \texttt{-1} & The number of self-consistent-field (SCF) steps for which to apply the  \\ & & preconditioning \cite{Corsetti2014}. If negative, the preconditioning is applied at all SCF steps \\
\texttt{OMM.PreconFirstStep} & \texttt{OMM.Precon} & \texttt{OMM.Precon} for the first MD step \\
\texttt{OMM.TPreconScale} & \texttt{10 Ry} & The scale $\tau_\mathrm{T}$ for the kinetic energy preconditioning (see Eq. (\ref{eq_precon})) \\
\texttt{OMM.Cholesky} & \texttt{false} & Whether to apply the Cholesky factorization \cite{Corsetti2014} \\
\hline
\end{tabular}
}
\label{table:input}
\end{table*}

\section{Tests}
\subsection{Computational details}
The test calculations have been carried out for single-layer boron nitride under periodic boundary conditions. Supercells of boron nitride from $12\times12$ to $96\times96$ with up to 18400 atoms are considered. The lattice constant is set at 2.48 \AA. The height of the simulation cell is 20 \AA. The calculations have been performed at the single $\Gamma$ point. The local density approximation \cite{Perdew1981}, norm-conserving Troullier-Martins \cite{Troullier1993} pseudopotentials and standard built-in DZP basis set \cite{Junquera2001} are used. The atomic orbitals are set to zero beyond the cutoff determined by the energy shift of 10 meV (cutoff radii 2.5 -- 4.5 \AA). The real-space grid is equivalent to the plane-wave cutoff energy of 100 Ry. The linear mixing scheme with a mixing parameter of 0.1 is applied to converge the ground state. The tolerance is 10$^{-4}$ for the density matrix and 10$^{-3}$ eV for the matrix elements of the Hamiltonian.  

To test performance of different approaches in MD simulations, several MD steps starting from the converged ground state have been computed (the ground state is converged previously with the same method as used for MD). The microcanonical ensemble with an initial temperature of 300 K is considered. The Verlet algorithm  \cite{Allen1987} with a time step of 1 fs is used. The Pulay mixing scheme \cite{Kresse1996} with a mixing parameter of 0.2  is applied during the MD simulations. 

The matrices involved in the calculations consist of equal blocks. For the DZP basis set, each boron and nitrogen atom has 13 basis functions, and hosts 3 or 2 wavefunctions depending on whether the unoccupied states are included into consideration or not, respectively. Therefore, the block size for the wavefunctions is usually chosen to be $b_\mathrm{WF}=6$ and for the basis functions $b_\mathrm{BF}=13$. The matrices are distributed on a 2D process grid. The cutoff radius for localized wavefunctions in typical calculations with sparse matrices is $R_\mathrm{C}=4$ \AA{}. The chemical potential for the Kim functional is $\eta =-5.5$ eV. CG iterations are performed until the difference of energies at consecutive CG iterations divided by the average energy at these iterations reaches 10$^{-9}$. The tests with the preconditioning for dense matrices have been carried out using the scale for the kinetic energy of $\tau_\mathrm{T}=10$ Ry (Ref. \cite{Corsetti2014}). 

\subsection{Results}

\begin{figure}
   \centering
 \includegraphics[width=\columnwidth]{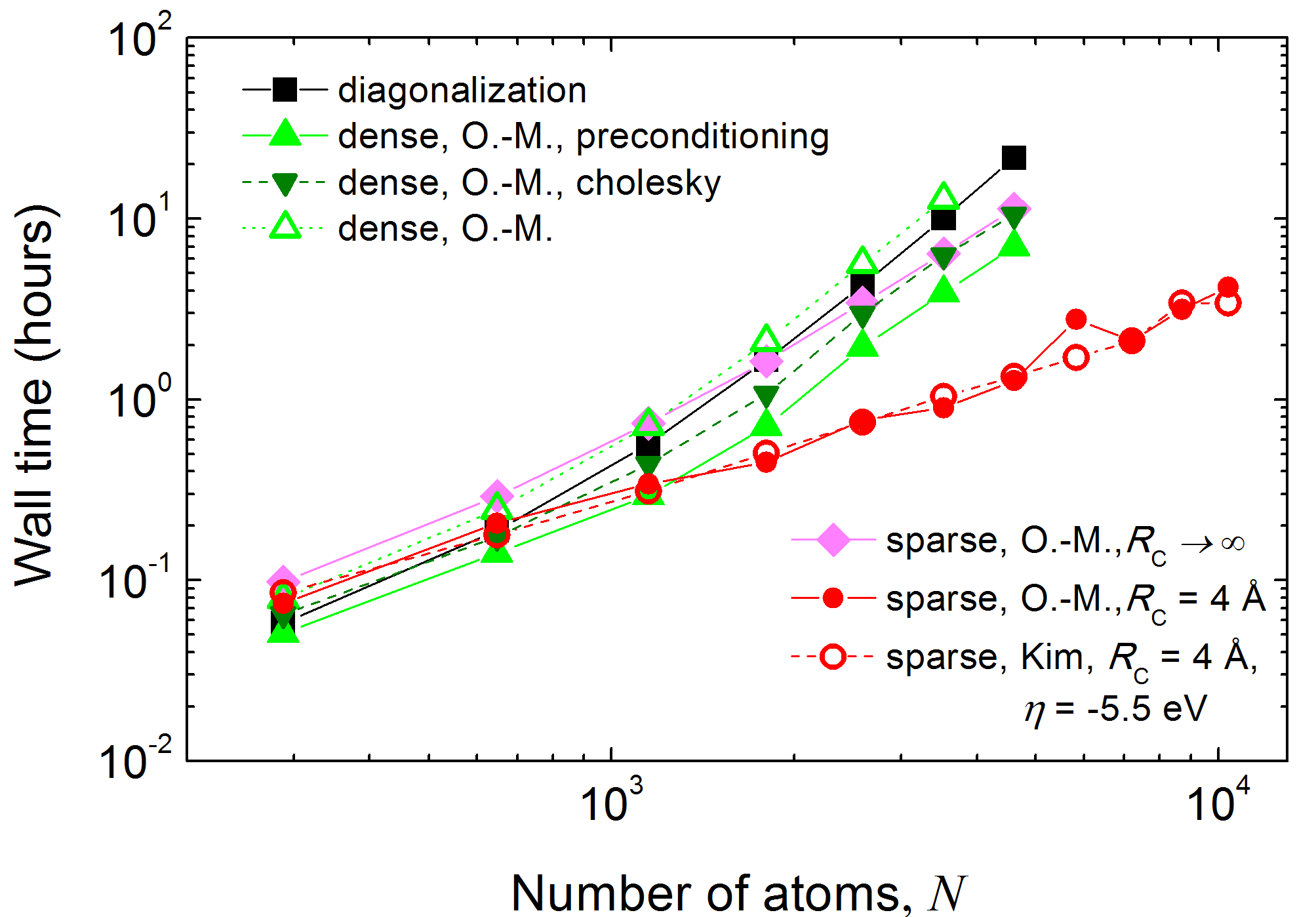}
   \caption{Wall time (in hours) for 4 MD steps for a single boron nitride (BN) layer computed using different approaches vs number $N$ of atoms in the system: (black squares) diagonalization, (green triangles up) OMM with dense matrices (using ScaLAPACK) and preconditioning using a kinetic-energy scale $\tau_\mathrm{T}=10$ Ry, (dark green triangles down) OMM with dense matrices with Cholesky factorization, (open green triangles up) plain OMM with dense matrices, (magenta diamonds) OMM with sparse matrices (using DBCSR) without wavefunction localization (wavefunction cutoff radius $R_\mathrm{C}\to\infty$), (red circles) Ordej\'on-Mauri functional with $R_\mathrm{C}=4$ \AA{} and (open red circles) Kim functional with $R_\mathrm{C}=4$ \AA{} and chemical potential $\eta =-5.5$ eV. In all the cases without wavefunction localization, the Ordej\'on-Mauri functional is considered. The calculations are performed on 96 CPU cores. A double-zeta polarized (DZP) basis set is used. The block size is $b_\mathrm{WF}=6$ for the wavefunctions and $b_\mathrm{BF}=13$ for the basis functions.} 
   \label{fig:size_methods}
\end{figure}

To compare the performance of diagonalization and OMM with dense and sparse matrices, we have performed test MD simulations for single-layer boron nitride (BN) in different sizes. Fig. \ref{fig:size_methods} demonstrates that the approaches in which the wavefunctions are not confined in space have much worse scaling with system size than the methods with localized wavefunctions within a cutoff radius $R_\mathrm{C}$. 
The scaling of the former approaches is close to cubic for large systems (exceeding 1000 atoms in our calculations). It should be noted, however, that for small systems (within 1000 atoms) the scaling is sub-cubic. The reason is that for such systems the solver contribution to the total time plotted in Fig. \ref{fig:size_methods} is comparable to the contributions of other parts of the code that have linear scaling with system size. Among the methods using dense matrices, OMM with applied preconditioning or Cholesky factorization, which improve convergence, shows a slightly better scaling compared to diagonalization or plain OMM. Also OMM using the DBCSR library with no localization of wavefunctions ($R_\mathrm{C}\to\infty$) clearly has a better scaling than OMM using ScaLAPACK. This is explained by the fact that the former, although having a dense coefficient matrix, still exploits the sparsity of the Hamiltonian and overlap.
 
In the range of system sizes considered, OMM with kinetic energy preconditioning is the fastest among the approaches without wavefunction localization, followed by OMM with the Cholesky factorization, diagonalization, and plain OMM (Fig. \ref{fig:size_methods}). The crossover between preconditioned dense OMM and the linear-scaling methods takes place for the system with about 1200 atoms. For the plain dense OMM and for diagonalization, the crossovers with linear-scaling methods occur earlier, at about 300 and 700 atoms, respectively.

\begin{figure}
   \centering
 \includegraphics[width=0.95\columnwidth]{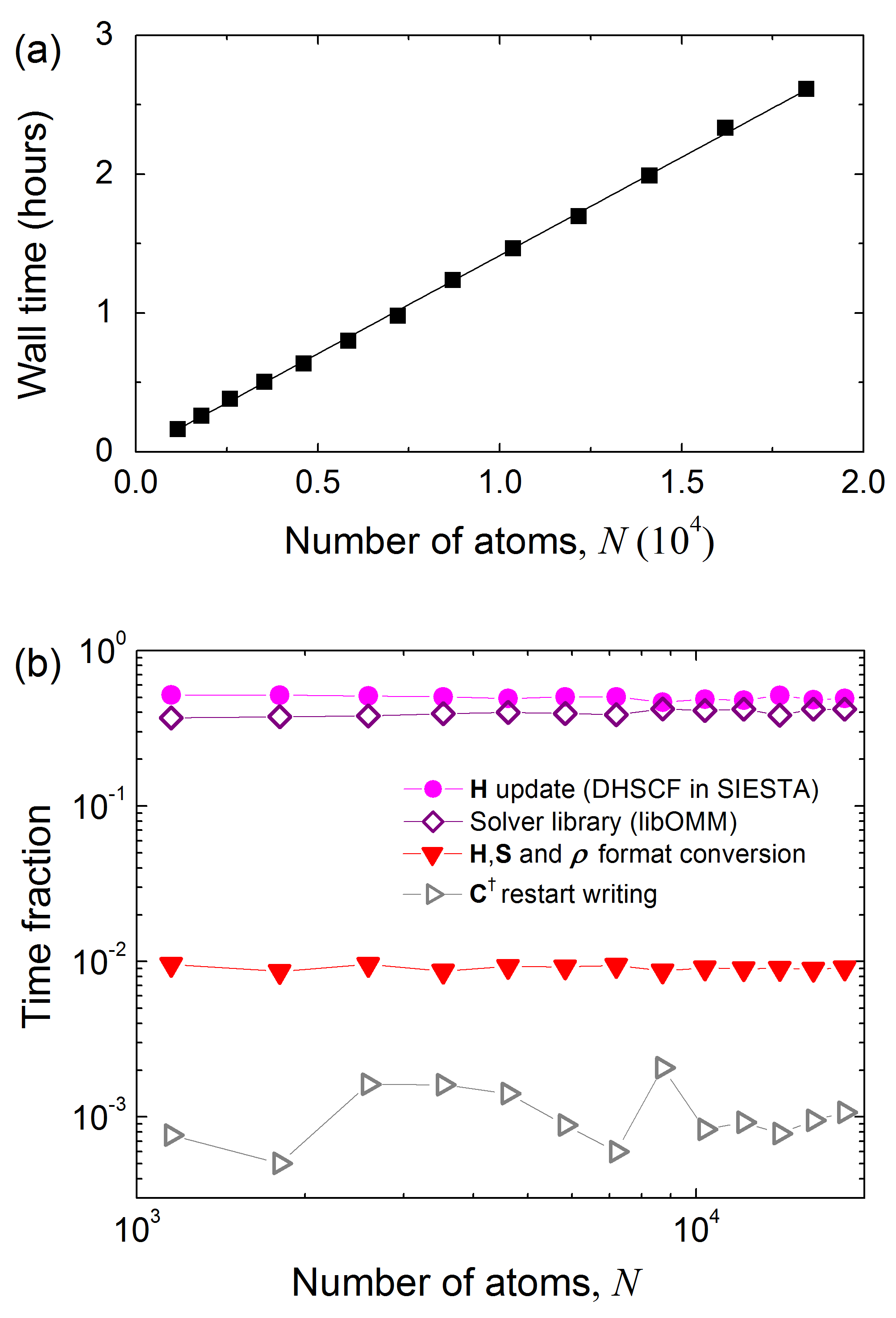}
   \caption{(a) Wall time (in hours) for 3 MD steps with 12 SCF iterations each for a single BN layer described using the Kim functional vs number $N$ of atoms (in $10^4$ atoms). A linear fit is shown by the solid line. (b) Relative contributions of the OMM solver subroutines to the total time vs $N$: (magenta circles) Hamiltonian update on a 3D grid (DHSCF subroutine in SIESTA \cite{Soler2002}), (open purple diamonds) solver library libOMM, (red triangles down) format conversion of $\mathbf{H}$, $\mathbf{S}$ and $\boldsymbol{\rho}$ and (open gray triangles right) writing restart for localized wavefunctions ($\mathbf{C}^\dagger$). The calculations are performed on 192 CPU cores for a DZP basis, $R_\mathrm{C}=4$ \AA{}, $\eta =-5.5$ eV,  $b_\mathrm{WF}=6$, and $b_\mathrm{BF}=13$.}
   \label{fig:size_subroutines}
\end{figure}

Our timings for single-layer BN have confirmed that the Ordej\'on-Mauri  and Kim approaches in which the wavefunctions are localized within a cutoff radius $R_\mathrm{C}$ show linear scaling with system size (Fig. \ref{fig:size_subroutines}a). The computational times corresponding to different parts of the solver (matrix conversion, libOMM library, initialization and update of the coefficient matrix, reading and writing of restart for localized wavefunctions) and other parts of the SIESTA code such as the subroutine for the Hamiltonian update called after the density matrix change at each SCF step (DHSCF), all do change linearly upon increasing the system size. As a result, relative contributions of different parts of the code do not depend on the system size (Fig. \ref{fig:size_subroutines}b). This is different from the cubic-scaling methods, in which the solver very early takes most of the computing time upon increasing the system size, since the rest of the code has linear scaling. It should also be noted that, for the system considered, the solver takes only 40--50\% of the computational time, comparable, for example, to the subroutine for the Hamiltonian update (DHSCF in SIESTA). Most of this time corresponds to the minimization of the energy functional given by Eq. (\ref{eq_OM}) performed by the solver library libOMM. The matrix format conversion takes only 0.5--1.0\% of the total time. Writing of the restart files for localized wavefunctions takes up to 0.3\% of the time, and initialization and update of the coefficient matrix take a negligible time within 0.01\%.

\begin{figure}
   \centering
 \includegraphics[width=\columnwidth]{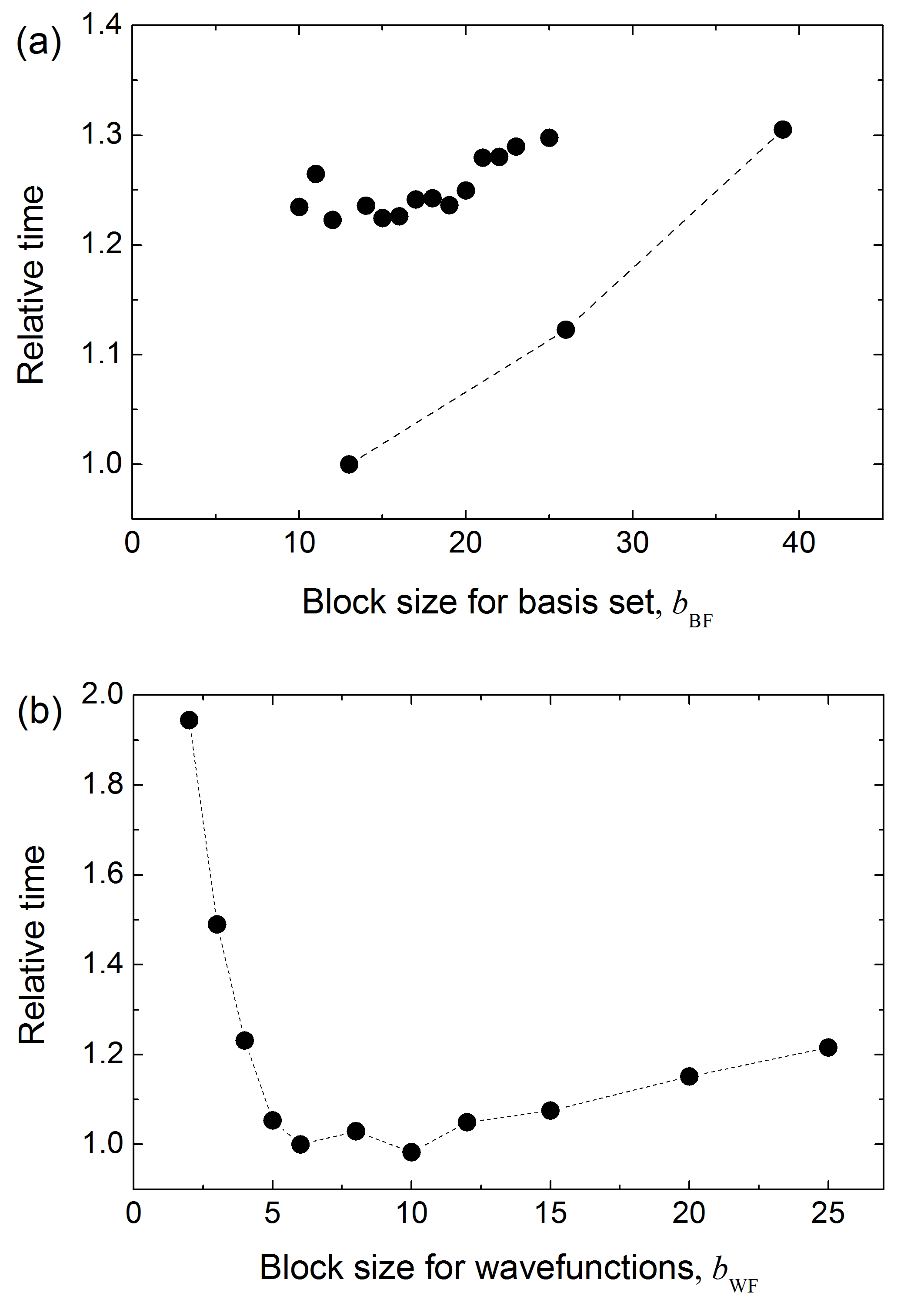}
   \caption{Relative time for the solver library libOMM during 4 MD steps for a single BN layer with a $60\times 60$ supercell (7200 atoms) using the Kim functional vs block size: (a) the block size for the basis set, $b_\mathrm{BF}$, is changed and the block size for the wavefunctions is kept as $b_\mathrm{WF}=6$ and (b) $b_\mathrm{WF}$ is changed and $b_\mathrm{BF}=13$. The relative time is given with respect to the result for $b_\mathrm{WF}=6$ and $b_\mathrm{BF}=13$. In panel (a), the dashed line is shown to guide the eye for the data obtained for $b_\mathrm{BF}$ divisible by the number of basis functions per atom. The calculations are performed on 192 CPU cores for a DZP basis, $R_\mathrm{C}=4$ \AA{}, $\eta =-5.5$ eV.} 
   \label{fig:blocksize}
\end{figure}

The dependence of computational time on block size for the Kim functional with DBCSR are presented in Fig. \ref{fig:blocksize}. In the case of the double-zeta polarized (DZP) basis set, each boron and nitrogen atom hosts 3 localized wavefunctions and 13 basis functions. Accordingly the computational time  drops significantly at block-size values $b_\mathrm{BF}$ for the basis functions divisible by 13 (Fig. \ref{fig:blocksize}a). For such block sizes, the computational time grows upon increasing the block size (note that the growth continues beyond the block sizes shown in Fig. \ref{fig:blocksize}a) and has the minimum at $b_\mathrm{BF}=13$. The wavefunction block-size $b_\mathrm{WF}$ dependence reaches the minimum at  $b_\mathrm{WF}=$ 6 -- 10. At small $b_\mathrm{WF}$, a fast growth of the computation time is observed. It can be attributed to an increase in the number of nonempty blocks considered upon decreasing the block size. At large $b_\mathrm{WF}$, the computational time also grows but at a slower rate. This dependence can be explained by increasing the number of matrix elements that are stored and explicitly considered in matrix operations. Therefore, we find optimal block sizes both for the wavefunctions and basis functions of the order of 10. Furthermore, chemical considerations can be exploited when dividing matrices into blocks. Still the optimal choice of block sizes for complex systems is not straightforward and requires further investigation \cite{CP2K}. 

\begin{figure}
   \centering
 \includegraphics[width=\columnwidth]{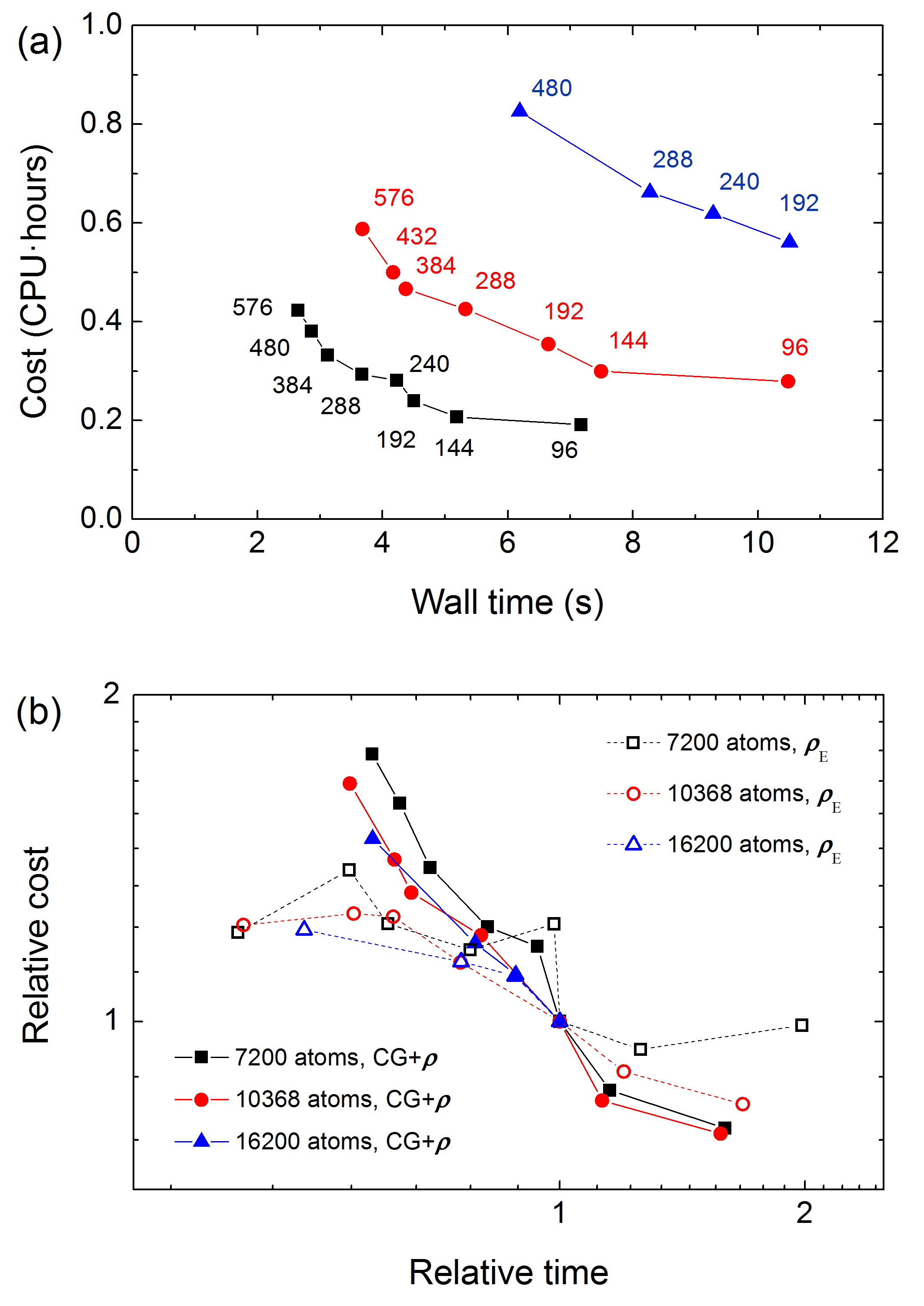}
   \caption{(a) Computational cost (in CPU$\cdot$hours) vs wall time (in s), for one call to the solver library libOMM including one conjugate gradient (CG) iteration and calculation of the density matrix $\boldsymbol{\rho}$, for different supercells of BN using the Kim functional: (black squares) $60\times 60$, (red circles) $72\times 72$ and (blue triangles) $90\times 90$ (7200, 10368 and 16200 atoms, respectively). The number of CPU cores used is indicated. (b) Relative computational cost vs relative time for the calls to the solver library including one conjugate gradient iteration and calculation of the density matrix $\boldsymbol{\rho}$ (closed symbols) or calculation of the energy density $\boldsymbol{\rho}_E$ (open symbols) for different supercells of BN. Relative values are given with respect to the results for 192 CPU cores. The calculations are for a DZP basis, $R_\mathrm{C}=4$ \AA{}, $\eta =-5.5$ eV, $b_\mathrm{WF}=6$, and $b_\mathrm{BF}=13$.}
   \label{fig:speedup}
\end{figure}

The CPU scaling of the libOMM solver library in calculations with sparse matrices using DBCSR is shown in Fig. \ref{fig:speedup}a. A similar CPU scaling is observed for systems of different size (Fig. \ref{fig:speedup}a), with different block and basis set sizes. The computational time decreases by a factor of about 2.5 upon doubling the computational cost. Such a speedup is observed for CG energy minimization and subsequent calculation of $\boldsymbol{\rho}$. It should be noted, however, that calls to libOMM for calculation of $\boldsymbol{\rho}_E$ involving only two matrix multiplication operations show much better CPU scaling. This can be appreciated from a twice steeper slope of computational cost versus computational time as compared to the calls for energy minimization and calculation of the density matrix (Fig. \ref{fig:speedup}b). It can, therefore, be expected that the solver parallelization might be further improved via proper code refactoring. The use of OpenMP, GPUs and the library for small matrix multiplication (LIBXSMM) \cite{LIBXSMM} are known to lead to a superior DBCSR performance \cite{Borstnik2014,CP2K}, which also requires investigation.

\begin{figure}
   \centering
 \includegraphics[width=\columnwidth]{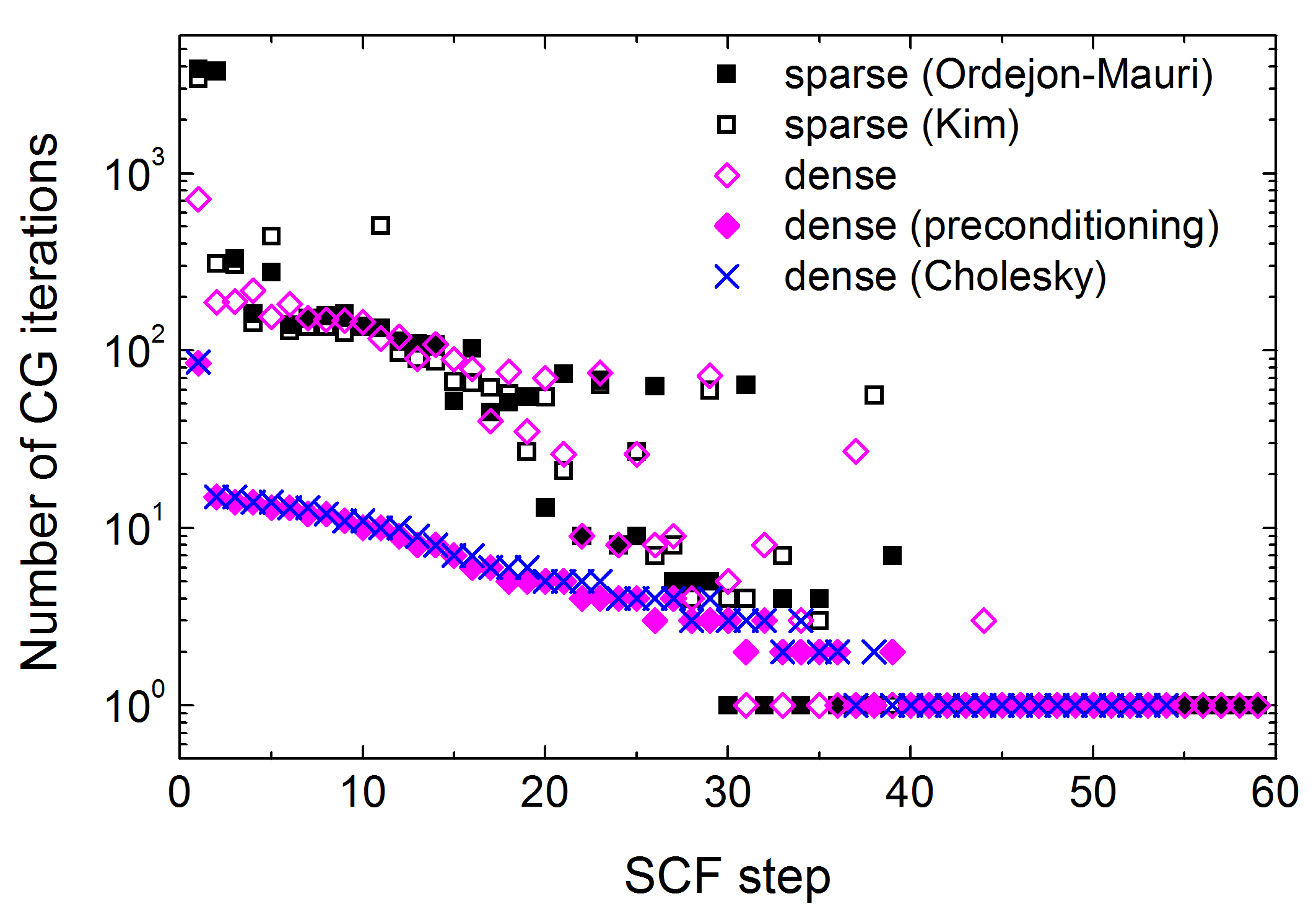}
   \caption{Number of conjugate-gradient (CG) iterations vs number of self-consistent-field (SCF) steps for the calculation of the ground state of a $12\times 12$ BN supercell (288 atoms) from scratch using OMM with sparse matrices: (closed black squares) Ordej\'on-Mauri and (open black squares) Kim functionals, and OMM with dense matrices using the Ordej\'on-Mauri functional: (open magenta diamonds) plain, (closed magenta diamonds) preconditioned with kinetic-energy scale of $\tau_\mathrm{T}=10$ Ry and (blue crosses) with the Cholesky factorization.  The calculations are performed on 96 CPU cores for a DZP basis, $R_\mathrm{C}=4$ \AA{}, $\eta =-5.5$ eV,  $b_\mathrm{WF}=6$, and $b_\mathrm{BF}=13$. Linear mixing with a mixing parameter of 0.1 is used.} 
   \label{fig:icg}
\end{figure}

\subsection{Recommendations for OMM solver use}
The new modular implementation of the OMM solver makes it easier to disentangle technical problems in e.g. parallelization from drawbacks of the OMM method itself. Here we present the first implementation of the solver utilizing external libraries that represents the starting point for further performance improvement and method polishing. Ways to improve the solver performance were mentioned in the previous subsection. We briefly discuss now the drawbacks of the OMM method and how they can be addressed. 

One of the most important methodological problems of the OMM approach is in the minimization, which can require a large number of CG iterations. As shown in Fig. \ref{fig:icg}, the first SCF iteration from scratch is rather costly both for the linear and cubic-scaling OMM. For the linear-scaling methods, the first SCF iteration can include thousands of CG steps, followed by tens of SCF iterations with hundreds of CG steps each. After that each SCF step needs just a few CG iterations, becoming very fast. It should be noted that except for the very first SCF iterations, the linear-scaling and plain cubic-scaling OMM require roughly same numbers of CG steps. However, kinetic energy preconditioning or Cholesky factorization significantly reduce the number of CG iterations required, with a considerable computational-time reduction (see also Fig. \ref{fig:size_methods}). Therefore, it is always recommended to use any of both ways to deal with kinetic energy ill-conditioning in dense OMM. The extension of these approaches to sparse matrices is not straightforward and requires further investigation.

Also starting from scratch, one can get into regions in parameter space where the energy functional does not have a minimum in the CG line minimization. To avoid this situation, we recommend using a small cutoff radius $R_\mathrm{C,ini}$ for the initial guess of wavefunctions both for linear and cubic-scaling OMM. It is also recommended to preconverge the ground state using a small linear-mixing parameter. Starting from as low as 0.01 can be required for very large systems. It can then be gradually increased to normal values of 0.1 -- 0.2. After getting close to the ground state, the use of other mixing schemes is possible. If the geometry of the system is far from the optimal one, a reduced step for geometry optimization may also be needed when starting. 

\begin{figure}
   \centering
 \includegraphics[width=\columnwidth]{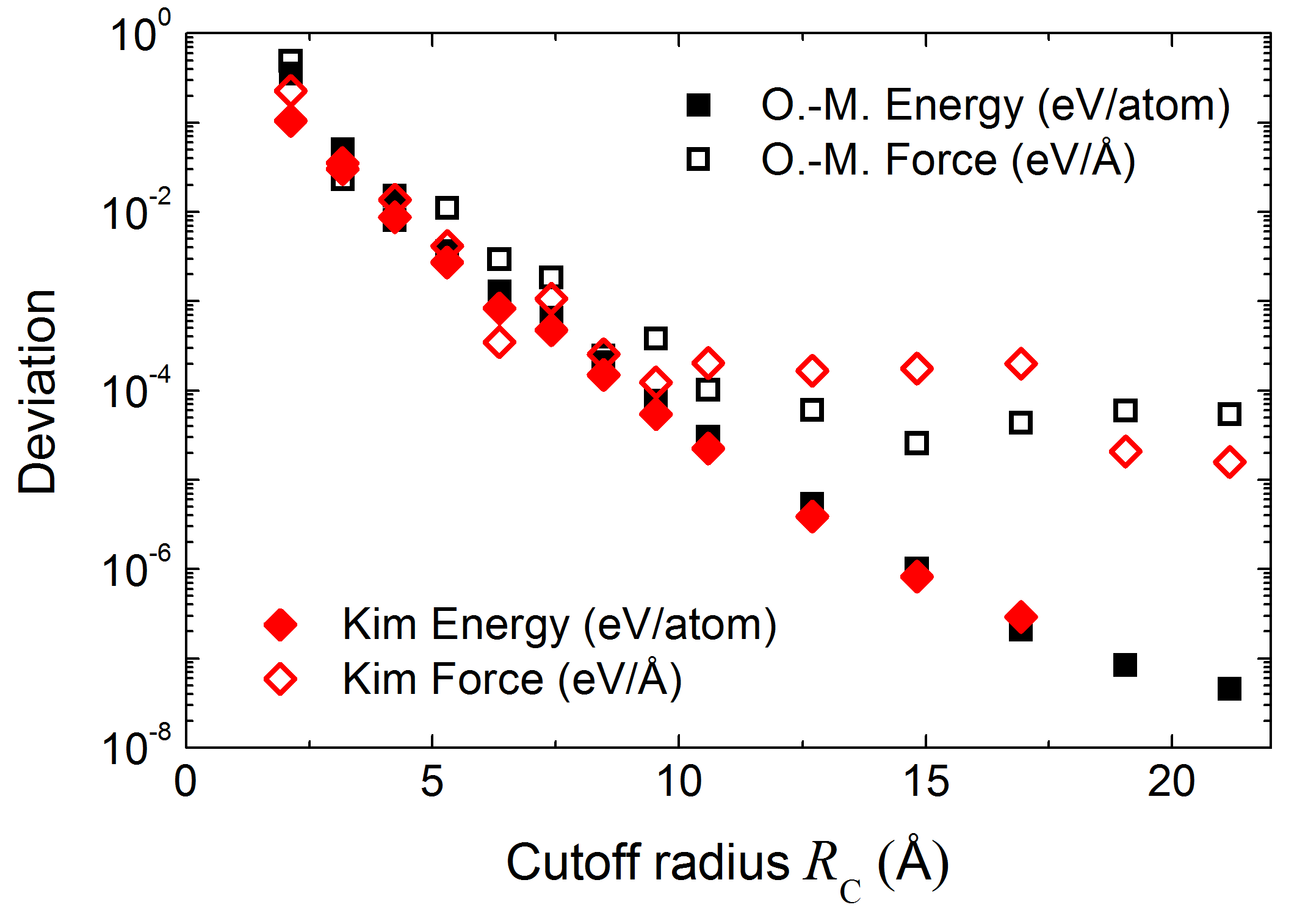}
   \caption{Deviations of energy (in eV/atom, closed symbols) and force (in eV/\AA, open symbols) for the $60\times 60$ supercell of boron nitride (7200 atoms) with atoms displaced by 0.05 \AA{} from their equilibrium positions from the results for the infinite cutoff radius for the wavefunctions $R_\mathrm{C}\to\infty$ vs cutoff radius $R_\mathrm{C}$ (in \AA):  (black squares) Ordej\'on-Mauri and (red diamonds) Kim methods. A DZP basis set is used. The chemical potential for the Kim method is $\eta =-5.5$ eV. The block size is $b_\mathrm{WF}=6$ for the localized wavefunctions and $b_\mathrm{BF}=13$ for the basis functions.} 
   \label{fig:rad_tol}
\end{figure}

In Fig. \ref{fig:rad_tol}, we address the accuracy of force and energy calculations  with the Ordej\'on-Mauri and Kim functionals for boron nitride. The deviation from the results for the wavefunctions without localization ($R_\mathrm{C}\to\infty$) is plotted for different cutoff radii $R_\mathrm{C}$. It is seen that for both of the functionals, the accuracy improves upon increasing the cutoff radius in a similar manner. The deviations of the energy and forces within 0.01 eV/atom and 0.02 eV/\AA{} are achieved already for the cutoff radius of $R_\mathrm{C}=4$ \AA. These results confirm that for insulating systems with a substantial band gap, it is sufficient to consider cutoff radii of several \AA{} \cite{Ordejon1995,Mauri1994}.  

The Ordej\'on-Mauri and Kim functionals were designed for insulating systems with a substantial band gap. For metals, a smearing function needs to be introduced. However, this is not easy since the information on individual Kohn-Sham eigenstates is missing in OMM. An idea for combining OMM with another method resolving eigenstates close to the Fermi level was proposed in Ref. \cite{Corsetti2014} but still requires exploration. Note that modeling of metallic systems requires a much more significant computational effort than modeling of insulators \cite{Goedecker1994, Goedecker1995, Mohr2018}. 

As for magnetic systems, the OMM calculations can be performed taking into account spin polarization. At each SCF step, the coefficient matrices for spin up and spin down are found sequentially. All the observations for non-spin-polarized systems discussed above still hold in this case.

\section{Conclusion}
We have demonstrated how modularization simplifies the implementation of new solvers in electronic structure codes by revising the OMM solver in the SIESTA code \cite{SIESTA,Ordejon1996,Soler2002,Sanchez-Portal1997,Garcia2020}. Matrix algebra operations and parallelization are efficiently handled via external libraries. In particular, the implementation benefits from two ESL \cite{ESL,Oliveira2020} libraries: libOMM\cite{LIBOMM,Corsetti2014,omm-bundle,Oliveira2020} and MatrixSwitch\cite{MS,MSDBCSR,omm-bundle,Oliveira2020}. The libOMM library is used to perform the minimization of the energy functional, while the MatrixSwitch library serves as an interface to low-level algebraic routines facilitating switching between different matrix formats. These libraries have been extended to make possible not only cubic-scaling but also linear-scaling OMM calculations for insulating systems with a substantial band gap. Now the energy functional minimization in libOMM can be carried out for sparse matrices with the DBCSR library \cite{DBCSR,Borstnik2014,CP2K}, in addition to dense matrices using ScaLAPACK \cite{SCALAPACK}. To facilitate incorporating libOMM into electronic structure codes based on atomic orbitals, MatrixSwitch has been also supplemented with subroutines for matrix format conversion and matrix reading and writing. The solver library libOMM can be easily further developed in the MatrixSwitch language for the implementation of new solvers. 

The extended MatrixSwitch and libOMM libraries available through ESL \cite{ESL,Oliveira2020} can be used for implementation of linear and cubic-scaling OMM approaches in other codes. The libraries can be used with different types of local basis sets. The only condition for achieving the linear-scaling behavior is that either the basis functions go to zero beyond some cutoff radius or the elements of the input matrices are filtered with respect to some tolerance to ensure that the matrices are sparse. Note that implementation of custom conversion routines is needed if the matrix format is different from the MS or SIESTA formats.

To test the performance of the new OMM and traditional diagonalization solvers available in SIESTA, large-scale calculations have been performed for a BN layer. When sparse matrices and localized wavefunctions are used, linear scaling with system size is achieved in practice, as expected. Matrix conversion, reading and writing of restart files, as well as initialization and update of the localized wavefunctions take a small fraction of the computational time. For the linear-scaling methods that fraction does not depend on system size. The cubic-scaling OMM with kinetic energy preconditioning performs best for small systems, even better than diagonalization. For plain OMM, diagonalization, and cubic-scaling OMM with kinetic energy preconditioning, the crossovers with linear-scaling methods are observed at about 300, 700 and 1200 atoms, respectively. The best performance for the linear-scaling OMM with sparse matrices is achieved when the wavefunctions and basis functions are divided into blocks of sizes around 10, taking into account the chemical structure. The OMM solver is MPI-parallelized. When using the DBCSR library \cite{DBCSR,Borstnik2014,CP2K} for algebraic operations with sparse matrices, the computational time decreases by a factor of 2.5 upon doubling the computational cost. It is expected that CPU scaling can be further improved via refactoring some operations in the libOMM library, using OpenMP and GPUs, etc.

To perform OMM calculations from scratch, it is recommended to start using a small linear-mixing parameter (down to 0.01), a small step for geometry optimization, and cutoff radii for the wavefunctions of a few \AA. For the cubic-scaling OMM, the convergence becomes much faster with kinetic energy preconditioning or Cholesky factorization. The extension of these approaches to sparse matrices demands further investigation. 

\section*{Data availability}
The data and relevant code for this research work are stored in GitLab: \url{https://gitlab.com/irina_lebedeva/siesta/-/tree/orderN} (SIESTA), \url{https://gitlab.com/ElectronicStructureLibrary/omm-bundle} (omm-bundle) and have been archived within the Zenodo repository: \url{https://doi.org/10.5281/zenodo.7781100} \cite{Lebedeva2023} (SIESTA), \url{https://doi.org/10.5281/zenodo.7781174} \cite{Lebedeva2023a} (MatrixSwitch and libOMM).
The raw data for tests have been archived within the Mendeley Data repository: \url{https://doi.org/10.17632/c8kz58bg5z.1} \cite{Lebedeva2022}.

\hfil
\section*{Acknowledgments}

The authors acknowledge the European Union MaX Center of Excellence (EU-H2020 Grant No. 824143), the Partnership for Advanced Computing in Europe (PRACE) for awarding us access to computational resources in Joliot-Curie at GENCI@CEA, France (EU-H2020 Grant No. 2019215186),
computational resources at Pirineus and the technical support provided by Consorci de Serveis Universitaris de Catalunya (RES grants No. FI-2022-1-0023 and FI-2022-2-0035) as well as technical and human support provided by IZO-SGI SGIker of the University of the Basque Country (UPV/EHU) and European funding (ERDF and ESF). ICN2 is supported by the Severo Ochoa program from Spanish MINECO (Grant No. CEX2021-001214-S) and by Generalitat de Catalunya (CERCA Programme). ICMAB is supported by the Severo Ochoa program from Spanish MICIU (Grant No. CEX2019-000917-S). PO acknowledges support by Spanish MICIU, AEI and EU FEDER (Grant No. PGC2018- 096955-B-C43). 
 AG acknowledges support by Spanish MICIU, AEI and EU FEDER (Grant No. PGC2018-096955-B-C44). P.O. and A.G. acknowledge support from Generalitat de Catalunya (Grant No. 2021SGR01519). EA acknowledges funding from Spanish MICINN through grant PID2019-107338RB- C61/AEI/10.13039/501100011033, as well as a Mar\'{\i}a de Maeztu award to Nanogune, Grant CEX2020-001038-M funded by MCIN/AEI/ 10.13039/501100011033. 
We also thank Dr. David L{\'{o}}pez-Dur{\'{a}}n for useful discussions and providing the introduction to the MatrixSwitch library.

\section*{Author contributions}
AG, EA and PO designed the project. IVL extended the codes and performed the calculations. All the authors discussed the results and commented on the manuscript.

\section*{Conflict of interest}

The authors declare no conflict of interest.
\end{document}